\documentclass[reprint,
 amsmath,amssymb,
 aps,superscriptaddress
]{revtex4-2}

\usepackage{graphicx}
\usepackage{dcolumn}
\usepackage{bm}
\usepackage{comment}

\begin{document}

\graphicspath{{./images}}

\preprint{APS/123-QED}

\title{Transport signatures of phase fluctuations in superconducting qubits}

\author{M. Wisne}
\affiliation{Department of Physics and Astronomy, Northwestern University, 2145 Sheridan Road, Evanston, IL 60208, USA}
\author{Y. Deng}
\affiliation{Department of Physics and Astronomy, Northwestern University, 2145 Sheridan Road, Evanston, IL 60208, USA}
\author{H. Cansizoglu}
\affiliation{Rigetti Computing, Berkeley, CA 94710, USA }
\author{C. Kopas}
\affiliation{Rigetti Computing, Berkeley, CA 94710, USA }
\author{J. Y. Mutus}
\affiliation{Rigetti Computing, Berkeley, CA 94710, USA }
\author{V. Chandrasekhar}
\affiliation{Department of Physics and Astronomy, Northwestern University, 2145 Sheridan Road, Evanston, IL 60208, USA}

\date{\today}

\begin{abstract}
Josephson junctions supply the nonlinear inductance element in superconducting qubits. In the widely used transmon configuration, where the junction is shunted by a large capacitor, the low charging energy minimizes the sensitivity of the qubit to charge noise while maintaining the necessary anharmonicity to qubit states.  We report here low-frequency transport measurements on small standalone junctions and identically fabricated capacitively-shunted junctions that show two distinct features normally attributed to small capacitance junctions near zero bias: reduced switching currents and prominent finite resistance associated with phase diffusion in the current-voltage characteristic. Our transport data reveals the existence of phase fluctuations in transmons arising from intrinsic junction capacitance.
\end{abstract}

\maketitle
Josephson junctions (JJs) form the heart of superconducting qubits, arguably the leading technology for quantum computing and quantum sensing.  In its simplest form, a JJ is a tunnel junction between two superconductors.  Two energy scales govern the physics of such a junction:  the Josephson energy $E_J$, which is proportional to the superconducting critical current $I_c$ of the junction, and the charging energy $E_C$, inversely proportional to the intrinsic capacitance $C$ of the junction. JJs have been studied now for more than four decades, first as essential elements of superconducting electronics and sensors such as superconducting quantum interference devices (SQUIDs) as well as a fundamental physics playground for macroscopic quantum effects \cite{voss_macroscopic_1981, devoret_measurements_1985, ambegaokar_quantum_1982, fisher_quantum_1985, clarke_macroscopic_1986}.  Adding to this rich history, JJs now form the critical component of superconducting quantum circuits. 

The first JJs were many tens of square microns in area, with correspondingly large critical currents $I_c$ and therefore large $E_J$. These JJs had relatively large intrinsic capacitances $C$ between electrodes and therefore low $E_C$, and thus were in the regime $E_C/E_J \ll 1$.  On the other hand, JJs in current manifestations of superconducting qubits are submicron in size, and $E_C$ is more comparable to $E_J$. Such junctions are the focus of this paper and are referred to as small-$C$ JJs from this point forward.  It has been known for some years that small-$C$ junctions are particularly sensitive to noise \cite{kautz_noise-affected_1990, biswas_effect_1970, caldeira_influence_1981}.  In order to decrease the susceptibility to charge noise in superconducting qubits, most qubits are fabricated in the so-called transmon configuration, where the JJ is shunted by an on-chip capacitor $C_S \gg C$.  Such transmon qubits show significant improvement in coherence times over unshunted JJs \cite{PhysRevX.9.041041, PRXQuantum.2.010339, PhysRevLett.130.267001}.

Here we report dc and low frequency electrical transport measurements on small-$C$ unshunted (``standalone'') JJs with $E_C/E_J \geq 1$ as well as shunted (``transmon'') JJs where the charging energy is several times reduced.  Somewhat surprisingly, we find that all devices show two features associated with small capacitance junctions subject to noise in their environment:  a finite zero-bias resistance in the nominally superconducting regime of the JJ current-voltage characteristic (IVC), and a switching current $I_{sw}$ out of the superconducting regime that is much lower than the expected critical current $I_c$ based on the measured characteristics of the JJ.  Most notably, we find that there is no significant difference between the dc transport characteristics of unshunted and shunted JJs with nominally identical junction areas and fabrication methods, contrary to theoretical predictions.  dc transport characterization on superconducting qubits is not routinely performed:  our results suggest that such measurements may be useful in optimizing qubit fabrication to increase coherence times.

\section{Background}
The dynamics of a JJ can be described by the quantum Langevin equation for a particle of mass $m=(\hbar/2e)^2 C$ in a potential $U(\phi)$ subject to a time- ($t$) dependent random force $F(t)$ \cite{FORD198829}
\begin{equation}
  m \ddot{\phi} + \int_0^\infty \mu(t - t') \phi(t') dt' + \frac{\partial U}{\partial \phi} = F(t).  
  \label{eq:langevin}
\end{equation}
where $\phi$ is the phase difference between the two superconductors forming the JJ.  $\mu(t)$ is the so-called memory function related to the autocorrelation of the random force $F(t)$ \cite{FORD198829}
\begin{align}
    \frac{1}{2} &<F(t) F(t') + F(t') F(t)> \nonumber \\
    =&\frac{1}{\pi} \int_0^\infty d \omega \; \Re[\tilde{\mu}(\omega)] \hbar \omega \coth(\hbar \omega/2 k_B T) \cos \omega(t - t'),
\label{eq:autocorrelation}
\end{align}
where $\tilde{\mu}(\omega)$ is the Fourier transform of $\mu(t)$, $k_B$ is Boltzmann's constant and $T$ is the temperature.  Equation \ref{eq:langevin} is identical to the equation for the Brownian motion of a particle of mass $m$ in a potential $U$ subject to the random force $F(t)$ \cite{elements}. This equation is a generalized form of the more familiar resistively and capacitively shunted junction (RCSJ) model for a JJ, where $U(\phi)$ is the tilted washboard potential $U(\phi) = -E_J \cos(\phi) - I \phi$, $E_J = (\hbar/2e) I_c$ being the Josephson energy and $I$ the current applied through the junction \cite{tinkham2004introduction}.  $I_c$ is given by the Ambegaokar-Baratoff relation \cite{ambegaokar_tunneling_1963}
\begin{equation}
    I_c R_n = \frac{\pi \Delta(T)}{2e} \tanh (\Delta(T)/2 k_B T)
    \label{en:ambegaokar-baratoff}
\end{equation}
where $R_n$ is the normal state resistance of the junction and $\Delta(T)$ the temperature-dependent gap.
The additional term $F(t)$ which is absent in the conventional RCSJ model represents the noise sources influencing the JJ.   Classically, one expects a particle with smaller mass to be more affected by a given random force $F(t)$. Indeed, since the mass of the fictitious phase particle is proportional to the intrinsic capacitance $C$, junctions with smaller $C$ tend to be more susceptible to noise. 

Let us consider some simple limits of Eqn. \ref{eq:langevin} to gain a better understanding of the dynamics of our phase particle and to define some parameters of the problem.  If we ignore $F(t)$ and its resulting dissipative term (the second term with the integral in Eqn. \ref{eq:langevin}) and consider the case with $I=0$, we can expand the wasboard potential about its minimum at $\phi=0$ to obtain
\begin{equation}
    m \ddot{\phi} + \frac{1}{2} m \omega_0^2 \phi^2 = 0
\end{equation}
which is the equation of a simple harmonic oscillator with the fundamental oscillation frequency given by $\hbar \omega_0 = \sqrt{4 e^2 E_J/C} = \sqrt{8 E_C E_J}$, where $E_C = e^2/2C$ is the single electron charging energy.  The quantum analog of this equation relevant for superconducting qubits corresponds to a harmonic oscillator with energy levels $E_n = (n+1/2) \hbar \omega_0$.  Considering the full anharmonic cosine potential results in the unequal level spacings required for qubit operation.  Note that to first order, $\omega_0$ is independent of the size of the junction; $I_c$ and $C$ scale with the area of the junction, so that the product $E_J E_C$ is independent of the area of the junction, while $E_C/E_J$ increases as the junction size decreases.

To include the effects of the random force and dissipation, it is useful to Fourier transform Eqn. \ref{eq:langevin}.  The response of the system can then be represented by $\tilde{\phi}(\omega) = \alpha(\omega) \tilde{F}(\omega)$, where $\tilde{\phi}$ and $\tilde{F}$ are the Fourier transforms of $\phi(t)$ and $F(t)$, respectively, and the response function $\alpha(\omega)$ is given by \cite{FORD198829}
\begin{equation}
    \alpha(\omega) = \frac{1}{-m\omega^2 + i\omega \tilde{\mu}(\omega) + m \omega_0^2}.
\end{equation}
The poles of $\alpha(\omega)$ give the excitation energies of the system.  If $\tilde{\mu}(\omega)=0$, we regain our earlier result for the excitation energy.  It would seem that including a finite $\tilde{\mu}(\omega)$ would introduce a decay time for the oscillation, or a finite lifetime for the state in the quantum limit.  However, this depends on the specific form of $\tilde{\mu}(\omega)$.  In our case, $\tilde{\mu}(\omega)$ can be related to the equivalent admittance of the environment coupled to the JJ, $\tilde{\mu}(\omega) = (\hbar/2e)^2 Y(\omega)$ \cite{joyez_josephson_1999}, where $Y(\omega)$ is the equivalent frequency dependent admittance considered connected in parallel to the junction.  Let us now consider the transmon configuration, where the JJ is shunted by a large capacitor $C_S$, $Y(\omega)= i\omega C_S$.  In this case, it is easy to show that the excitation energy is then modified to $\hbar \omega_0 = \sqrt{8E_J E_C'}$, where $E_C' = e^2/2(C + C_S)$, as expected.

More generally, the environmental coupling to the JJ, which includes the impedances of the measurement electronics, the wiring in the cryostat, etc., may include an ohmic component that will give rise to dissipation.  In the context of the RCSJ model, dissipation is embedded in a shunt resistance $R$ in parallel to the JJ, so that $Y(\omega)= 1/R$.  While the exact origin of $R$ seems to be a matter of debate, it is commonly thought that $R$ is the resistance associated with quasiparticle transport across the junction \cite{orr} combined with any ohmic contributions from the environment such as the source impedance of the electronics connected to the JJ \cite{kautz_noise-affected_1990, joyez_josephson_1999, hu_low-voltage_1992}.  Associated with $R$ are fluctuations in the current $\Delta I(t)$ attributed to Johnson noise and are described by the fluctuating random force $F(t)$ in the Langevin equation (Eqn. \ref{eq:langevin}). 

In principle, the Josephson phase particle, treated as a quantum variable, can thermally activate or tunnel through the barrier between adjacent minima of the tilted cosine potential. This is more likely for currents $I$ approaching $I_c$, where the potential barrier is substantially reduced.  In underdamped junctions where motion of the phase particle does not dissipate much energy, there is a high likelihood that once the particle thermally activates or tunnels out of the well, it will continue to time evolve in the so-called free-running state, leading to a finite voltage across the JJ \cite{tinkham2004introduction}. At this point, the junction is said to have entered the quasiparticle regime, where quasiparticle transport across the junction results in ohmic dissipation.  Consequently, the switching current $I_{sw}$ at which the JJ transitions to the finite voltage state is typically less than the critical current $I_c$ expected from the Ambegaokar-Baratoff relation Eqn. \ref{en:ambegaokar-baratoff}.  Thus, one expects small-$C$ junctions to be more susceptible to early switching, such that the ratio $I_{sw}/I_c$ decreases with increasing $E_C/E_J$, a trend that has been recently verified experimentally \cite{lu_phase_2023}.  

Relevant for the operation of qubits, at $I\sim 0$ there is a finite tunneling probability between adjacent minima of the periodic washboard potential. If there exists an available state in a neighboring well, the particle can experience a phase jump of $2\pi$ without entering the full running state. Over time, many successive phase jumpss results in phase diffusion associated with a finite voltage for currents $I<I_{sw}$ in the nominally superconducting regime of the JJ.  The corresponding ``zero-bias'' resistance $R_0$ has been observed for many years in small-$C$ JJs, beginning with Ref. \cite{martinis_1985}.  Theoretically, one expects $R_0$ to increase super-exponentially with increasing $E_C/E_J$, a result that has also been verified experimentally \cite{iansiti_charging_1987, iansiti_crossover_1988, martinis_classical_1989, joyez_josephson_1999}.

With the amount of attention JJs have garnered in the last half-century, one may assume our understanding of single junction dynamics to be complete. However, in anticipation of the following discussion, we reiterate two important pieces of information. First, we point out that phase tunneling is not exclusive to junctions where $E_C \sim E_J$. The extent to which these phase fluctuations negatively influence qubit operation is unknown. Second, we emphasize the incomplete nature of our understanding of how environmental impedance influences such JJ phase dynamics, especially in the context of the high-impedance environment of superconducting qubits. A goal of this paper is to draw attention to features inherent to single Josephson junctions that could pose challenges to superconducting qubit performance. 

\section{Device fabrication and measurement details}
The Al/AlO\textsubscript{x}/Al junctions intended for superconducting qubits included in this study were fabricated using electron-beam lithography without a suspended bridge \cite{bridgeless}.  Details of the fabrication process can be found in the Supplementary Materials, but we point out key features relevant to our discussion here.  The areas of the junctions were defined by a fixed width of 240 nm and lengths varying from 260 to 1510 nm, yielding nominal overlap areas between 0.06 and 0.36 $\mu\text{m}^2$.  Two different sets of devices with these junction areas were fabricated and measured:  single ``standalone'' junctions with Al bonding pads and single junctions that were connected to Nb shunting capacitors in a transmon configuration.  The intrinsic capacitance of the JJs is estimated to be approximately 4-22 fF, depending on junction size, while the additional shunt capacitance $C_S$ in the transmon configuration is calculated to be $\sim$100 fF by electrostatic simulation prior to fabrication.

To address concerns over intrinsic defects associated with this particular fabrication process, we also fabricated and measured standalone junctions using the familiar Dolan shadow evaporation technique \cite{dolan_offset_1977} to compare with the bridgeless junctions. A double angle of $\pm 15 ^{\circ}$ in PMMA/MMA bilayer resist with 100 mTorr of oxygen introduced between Al depositions was used to create Dolan junctions with similar areas to the bridgeless process.

Transport measurements were performed in an Oxford MX100 dilution refrigerator at temperatures down to $\sim$20 mK using a four-terminal measurement configuration.  For the dc IV measurements, the dc current was applied using a custom room temperature current source with a nominal output impedance of $\sim10^{13} \; \Omega$, while the voltage across the JJ was measured using a custom room temperature voltage preamp with an input impedance of $10^9 \; \Omega$.  For the differential resistance ($dV/dI$)  measurements, a low frequency (10s of Hz) ac current was superposed on the dc current, and the resulting ac signal was demodulated using a PAR124 lock-in amplifier.  In order to capture the full extent of the features observed at low bias, ac currents of order 25 pA rms were used.  It is important to note that unlike some other works on small capacitance JJs, no large on-chip or low-temperature resistors were used to isolate the JJ.  This means that no offset voltages due to the Coulomb blockade of charge were expected in the IV characteristics, a fact that was confirmed by applying a large magnetic field to suppress superconductivity.  With such a field applied, the IV characteristic appeared perfectly linear and antisymmetric with respect current drive, with a slight increase in slope near $I=0$ associated with the dynamical Coulomb blockade \cite{dcb} discernible only in the $dV/dI$ traces (see Supplementary Materials).

\section{Experimental results}

Figure 1 shows a typical IVC for a standalone JJ in the regime $E_C/E_J \sim 1/10$ with area $\sim0.25 \; \mu\text{m}^2$. Since several features in this IVC are critical to our discussion of transmons, we explain them in detail in the following text. 

\begin{figure}[h!]
    \centering
    \includegraphics[width = 8.5 cm]{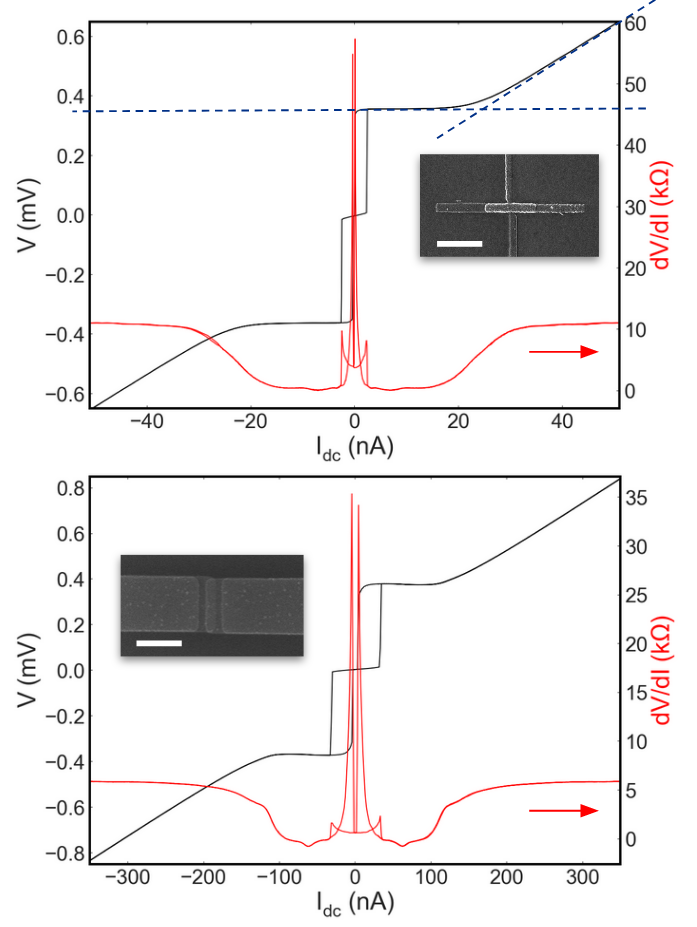}
    \caption{Simultaneously measured IVC and \textit{dV/dI} on small standalone junctions fabricated using a bridgeless evaporation technique (top) and traditional Dolan shadow evaporation (bottom) at 20 mK. Inset: scanning electron microscope images of junctions. Scale bar denotes $1\;  \mu\text{m}$.}
    \label{fig:singlejj}
\end{figure}

Before cooling superconducting qubits in a dilution refrigerator, $E_J$ is often estimated from the measured junction resistance at room temperature and the assumed gap $\Delta$ of aluminum. However, one can also directly extract $E_J$ from the IVC at low temperature using Eqn. \ref{en:ambegaokar-baratoff} where $\Delta(T\rightarrow0)$ and $R_n$ are measured quantities. Consider, for example, the bridgeless junction IVC in Fig. 1. The dotted lines indicate the slope at high bias ($R_n$) and the onset of the quasiparticle branch ($V = 2\Delta/e$) used to calculate $E_J$ and $I_c$ from Eqn. \ref{en:ambegaokar-baratoff}. With $R_n = 11.42 \; \Omega$ and $\Delta = 188 \; \mu{eV}$, the Josephson energy $E_J$ of the bridgeless junction is found to be $55\; \mu \text{eV}$. Since tunnel junction resistances increases at low temperature and small differences in aluminum disorder vary $\Delta$, we suspect that, when possible, $E_J$ extracted from the IVC provides a more accurate description for junction energy compared to room temperature approximations. 

While $E_C$ cannot be directly determined from the IVC, an estimate of $C$ given the dielectric constant of aluminum oxide and junction size yields an approximate charging energy $E_C = 4 \; \mu eV$ \cite{C_estimate}. For the bridgeless standalone junction in Fig. 1, then, $E_C/E_J \sim 1/14$ and $I_C = 27 \;\text{nA}$. Notably, however, the current at which the junction is \textit{measured} to switch occurs at a much smaller value, the average being $I_{sw} \sim 3 \; \text{nA}$ for this particular sample. The switching current is also reduced in the Dolan junction, though the ratio $I_c/I_{sw}$ is larger, agreeing with trends in the literature for $E_J$ and $I_{sw}$ \cite{lu_phase_2023}. 

\begin{figure*}[ht]
    \centering
    \includegraphics[width = \textwidth]{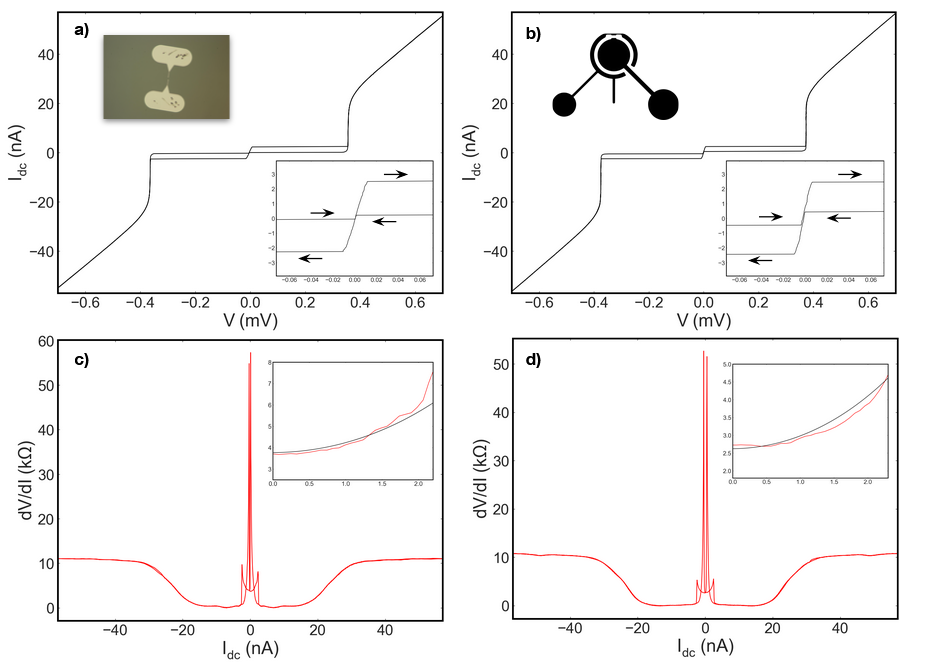}
    \caption{Transport measurements of standalone JJ (left) and junction shunted by a 100 fF on-chip capacitor (right). Standalone JJ IVC (a) with optical image of entire device (top inset) and transmon JJ IVC (b) with schematic showing circular shunt capacitor and protruding wirebond pads (top inset). Arrows show direction of bias current sweep magnified near zero bias. Standalone and transmon JJ $dV/dI$ (c) and (d), respectively, with nonlinear $R_0$ around zero bias (insets) magnified to show experiment (red) and theory (black).}
    \label{fig:enter-label}
\end{figure*}

One may observe switching currents $I_{sw} \ll I_c$ for a several reasons, e.g. inadvertent electron heating due to the biasing circuit. At millikelvin temperature, however, we expect quantum phase fluctuations to dictate the switching behavior. Since $I_{sw}$ is the maximum current carried in the phase diffusion regime of the IVC before spontaneously transitioning into a higher voltage state, it gives the experimentalist an idea of the degree of environmental noise coupled to the system. $I_{sw}$ is a stochastic variable that depends weakly on dc slew rate, or how quickly the potential is tilted; a slower tilt allows more time for probabilistic particle escape, which causes early activation into the running state. Varying the slew rate from 1 aA/s to 1 $\mu$A/s extends the average $I_{sw}$ from 2 nA to 5 nA (see Supplementary Material for further details).  For the purposes of this experiment, we take $I_{sw}$ to be the peak value of the switching distribution. 

Though not critical to the aim of this paper, we point out the signature feature of underdamped junctions is hysteresis in the IVC, as seen in Fig. 1. The potential tilt required to relocalize an underdamped phase particle is steeper than the tilt at $I = I_{sw}$. The so-called retrapping current $I_r$ is therefore a smaller fraction of $I_c$ in underdamped junctions. The difference between $I_{sw}$ and $I_r$ appears as hysteretic behavior in both directions of current sweep of the IVC.

The finite slope in Fig. 1 where $I<I_{sw}$ depicts phase diffusion of a continuously tunneling Josephson particle described earlier in the text. Phase diffusion in small JJs has been experimentally observed for decades \cite{iansiti_charging_1987, iansiti_crossover_1988, kautz_noise-affected_1990, martinis_classical_1989} with various theoretical explanations \cite{biswas_effect_1970, IZ, hu_low-voltage_1992, caldeira_influence_1981}. In the present case, the zero-bias resistances $R_0 = dV/dI$ as $I \rightarrow 0$ agrees with past experiments on underdamped JJs of similar sizes \cite{lu_phase_2023}. Moreover, we also observe non-linear dependence of $R_0$ with current bias $I$ which, until very recently, lacked a theoretical model \cite{chandrasekhar} that we briefly summarize here. When current across the junction is increased, there are two competing phenomena describing the resistance near zero bias: exponentially increasing tunneling probability from the diminishing height of the cosine potential and decreasing ground state energy of the phase particle. These combine to form the nonlinearity of $R_0(I)$ observed in Fig. \ref{fig:singlejj}. 

It has been suspected that the measurement circuit influences $R_0$ much the same way that $R$ in the RCSJ model depends on environmental impedance. However, replacing the high-impedance current source with a pseudo-current source setup utilizing a voltage-biased $100 \; \text{k}\Omega$  source resistor in series with the JJ had no effect on the phase diffusion slope of the IVC. Additionally, Ref. \cite{chandrasekhar} found that, compared to the ratio $E_C/E_J$, environmental admittance $Y(\omega)$ has little effect on $R_0$.

As $I \rightarrow 0$, $R_0 = 3.63$ k$\Omega$ in the bridgeless junction and $R_0 = 600$ k$\Omega$ in the Dolan junction. As previously discussed, zero-bias resistance $R_0$ represents phase fluctuations or the continuous diffusion of phase through the cosine potential. $R_0$ on the order of $10^3 \; \Omega$ at milliKelvin temperature is attributed predominately to phase tunneling since the crossover temperature from tunneling to thermal activation \cite{BLACKBURN20161} greatly exceeds the temperatures used in this experiment. We assume low electron temperature for several reasons. First, milliKelvin temperature is substantiated in the sharpness of the quasiparticle tunneling regime of the IVC. Deliberately increasing temperature results in Fermi-function-like rounding at $V=2\Delta/e$ as $T \rightarrow \;$300 mK. The onset of the quasiparticle branchs in Fig. 1 are considered substantially sharper. Second, we note that the region of differential negative resistance observed in the $dV/dI$ disappears in warm environments. This region is associated with the time-correlated tunneling of single Cooper pairs, with the back-bending region in the IVC defined as the so-called ``Bloch nose'' \cite{corlevithesis}. The Bloch nose feature is washed out at higher temperature, similar to switching current.  

\section{Discussion}
Transport characterization was performed on over 20 capacitively shunted and unshunted junctions fabricated using bridgeless evaporation and Dolan double angle techniques. A main finding of this paper is the surprising similarity between the transport characteristics of single standalone junctions and identically fabricated junctions shunted with a large on-chip capacitor. As previously mentioned, the Langevin model predicts that as the effective device capacitance is increased with respect to constant $E_J$, one expects a decreased susceptibility to the random force $F(t)$ that results in reduced switching currents out of a local potential minimum. According to our measurements, however, there is little difference in $I_{sw}$ for capacitively shunted and unshunted junctions. Figure 2 shows transport measurements performed on two junctions with nearly identical $E_J$ but different $E_C$, one in a transmon configuration, the other standalone. Despite $E_C/E_J$ ranging from $1/10$ to $1/50$, switching currents in both devices agree very well with previous data for $E_C \sim E_J > k_bT$ \cite{lu_phase_2023}. This is an unexpected result. dc transport theory developed for JJs where $E_C/E_J \ll 1$ -- under which one may traditionally categorize transmons -- does not adequately explain the reduced switching current observed in these junctions. It is suggested that delta-correlated Johnson noise of the biasing electronics generates incoherent phase slips which drive the junction into the voltage state $V = 2\Delta/e$ \cite{IZ}. However, the high-impedance current source used in this study lie outside this regime and are not considered to influence switching currents. This heavily implies that the intrinsic capacitance of the junction, not the shunt, dictates phase dynamics. It appears that while $E_C/E_J$ determines qubit resonance and anharmonicity, it is not quite as straightforward a measure of the degree of phase fluctuations in capacitively shunted junctions.

\begin{figure}[h]
    \centering
    \includegraphics[width = 8.5 cm]{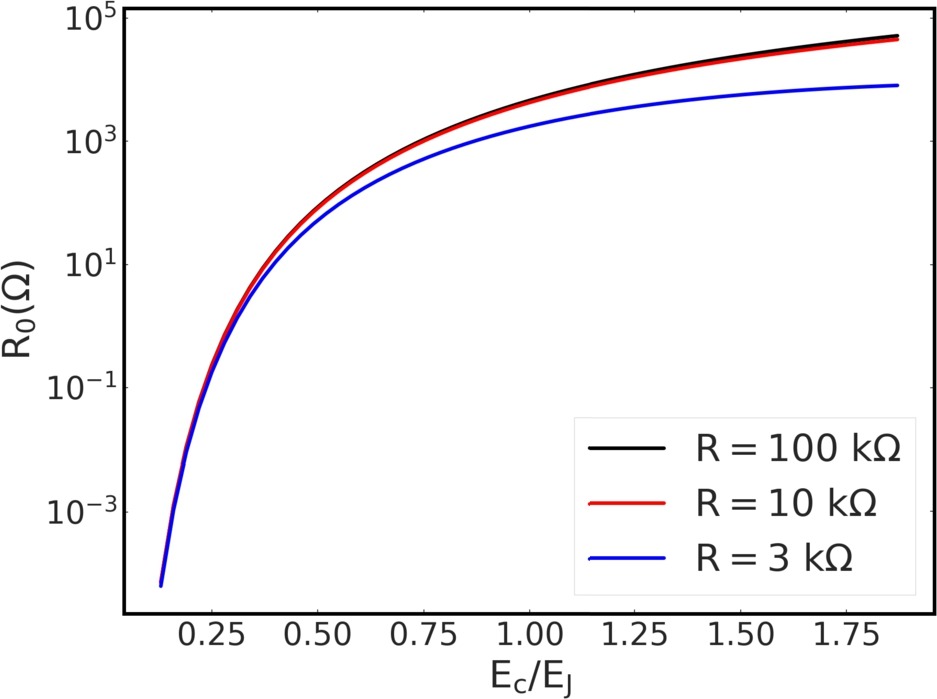}
    \caption{Simulated zero-bias resistance $R_0$ as a function of $E_C/E_J$ using quantum Langevin equation from Ref. \cite{chandrasekhar}. $10^3 \; \Omega$ resistance corresponds to unity $E_C/E_J$ regardless of shunt resistance $R$. The value of $E_J$ is fixed and corresponds to $I_C = 30$ nA.}
    \label{comp}
\end{figure}

Additionally, transmon phase fluctuations mirror that of the standalone junction in the zero-bias resistance $R_0$. The differential resistances for the shunted and unshunted JJs in Fig. 2 closely resemble each other. Extracted from the $dV/dI$, we find $R_n = 11.42$ for the standalone junction and $R_n = 11.06 \; \text{k}\Omega$ for the transmon, implying a similarity in junction area between the two. As $I \rightarrow 0$, $R_0 =$ 2.78 k$\Omega$ for the transmon and $R_0 =$ 3.63 k$\Omega$ for the standalone JJ. These numbers are typical for junctions of this size measured in this study. The fact that $R_0$ does not vanish in the capacitively shunted junction is somewhat surprising. According to Ref. \cite{chandrasekhar}, $R_0$ depends exponentially on $E_C/E_J$ for junctions with $E_J \approx 50 \; \mu\text{eV}$. We would expect, then, a drastic decrease in $R_0$ with decreasing $E_C/E_J$. However, equating the measured $R_0$ to the current-dependent tunneling rate suggests the effective ratio $E_C/E_J$ is very nearly the same between the shunted and unshunted junction. In fact, according to calculations using the quantum Langevin equation and dissipative quantum tunneling for underdamped junctions \cite{hu_low-voltage_1992, chandrasekhar}, the magnitude of $R_0$ for the standalone and transmon JJs correspond to $E_C/E_J \sim 1$. See Fig. 3. The energy ratio $E_C/E_J$ calculated from the IVC and junction area is incommensurate with the measured zero-bias resistance; one expects $R_0 \ll 1$ k$\Omega$ for junctions with $E_C/E_J < 1$. In single junctions, $R_0$, which describes quantum tunneling, has been to found to scale inversely to $E_J$ \cite{watanabe}. To our knowledge, however, measurements on small-$C$ single junctions with $R_0 \sim 1$ k$\Omega$ and well-defined switching current have never been reported in the literature. Puzzling still is the robust nature of $R_0$, which varies little between capacitively shunted and unshunted junctions and implies the external shunt capacitor has little effect on the presence of phase fluctuations.

The authors note an absence of transport measurements performed on single Josephson junctions destined for superconducting qubits in the literature. Recent devices utilizing SQUIDs, JJ arrays, on-chip impedances and cold resistors tend to focus on ultra-small junctions where $E_C/E_J \leq 1$. With the data reported in this paper, we emphasize isolated, inherent Josephson phase dynamics that lie at the heart of the qubit. We believe dc transport is a powerful tool in the growing repertoire of qubit characterization.  

In summary, we observe two features characteristic of small-capacitance JJs -- reduced switching current and zero-bias resistance -- in the transmon configuration, suggesting that phase fluctuations, which signify intrinsic noise, ignore the shunt capacitance and could fundamentally limit the performance of superconducting qubits. Our findings suggest the intrinsic capacitance of the junction matters more than previously assumed, and might motivate work on qubits that incorporate the shunt into the junction itself, as in the merged-element transmon design \cite{merged}, which could potentially mitigate phase fluctuations. 

This work is supported by the U.S. Department of Energy, Office of Science, National Quantum Information Science Research Centers, Superconducting Quantum Materials and Systems Center (SQMS) under Contract No. DEAC02-07CH11359.

\bibliography{JJ-dissipation}

\end{document}